\documentclass[a4paper,english]{lipics}

\usepackage{microtype}

\title{Games with Delays -- A Frankenstein Approach}
\author{Dietmar Berwanger}
\author{Marie van den Bogaard}
\affil{LSV, CNRS \& ENS Cachan, France}
\Copyright[nc-nd]%choose "nd" or "nc-nd"
          {Dietmar Berwanger, Marie van den Bogaard}

% mandatory: Please provide 1-5 keywords
\serieslogo{}%please provide filename (without suffix)
\volumeinfo%(easychair interface)
  {Editors}% editors
  {1}% number of editors: 1, 2, ....
  {CSL}%event
  {1}% volume
  {1}% issue
  {1}% starting page number
\EventShortName{}
\usepackage{amssymb,amsmath,xspace,enumerate}
\usepackage{tikz}
\usetikzlibrary{shapes,arrows}

\renewcommand{\epsilon}{\varepsilon}
\renewcommand{\phi}{\varphi}
\renewcommand{\theta}{\vartheta}

\usepackage{enum}

\usepackage[algoruled]{algorithm2e}
\DontPrintSemicolon

\def\qed{\ifmmode\squareforqed\else{\unskip\nobreak\hfil
\penalty50\hskip1em\null\nobreak\hfil\squareforqed
\parfillskip=0pt\finalhyphendemerits=0\endgraf}\fi}

\usepackage{url}

\newcommand*{\VA}{V \kern-2pt A}
\newcommand*{\mean}{\mathrm{mean}\text{-}\mathrm{payoff}}
\newcommand*{\LimSup}{\mathrm{limsup}}
\newcommand*{\parity}{\mathrm{parity}}
\newcommand{\takeout}[1]{}

\begin{document}

\title{Games with Delays -- A Frankenstein Approach}

\maketitle
\thispagestyle{empty}

\begin{abstract}
We investigate infinite games on finite graphs where the information flow is
perturbed by nondeterministic signalling delays. It is known that such
perturbations make synthesis problems virtually unsolvable, in the
general case.  On the classical model where signals are attached to
states, tractable cases are rare and difficult to identify. 

Here, we propose a model where signals are detached from control
states, and we identify a subclass on which equilibrium outcomes can
be preserved, even if signals are delivered with a delay that is 
finitely bounded. 
To offset the perturbation, our solution procedure combines
responses from
a collection of virtual plays following an equilibrium strategy in the
instant-signalling game
to synthesise, in a Frankenstein manner,
an equivalent equilibrium strategy for the delayed-signalling
game. 
\end{abstract}

\section{Introduction}

Appropriate behaviour of an interactive system component
often depends on events generated by other components.
The ideal situation, in which perfect information is available across
components, occurs rarely in practice -- typically a component 
only receives signals more or less correlated with the actual events. 
Apart of imperfect signals generated by the system components, 
there are multiple other sources of  
uncertainty, due to
actions of the system environment or to unreliable
behaviour of the infrastructure connecting the components: 
For instance, communication channels
may delay or lose signals, 
or deliver them in a different order than they were emitted.
Coordinating components with such imperfect information
to guarantee optimal system runs is a significant, but 
computationally challenging problem, in particular 
when the interaction is of infinite duration. 
It appears worthwhile to study the different sources
of uncertainty in separation rather than as a global phenomenon, 
to understand their computational impact on the synthesis of 
multi-component systems.

In this paper, we consider interactive systems modelled by concurrent
games among multiple players with imperfect information over finite
state-transition systems, or labelled graphs. Each state is associated
to a stage game in which the players choose simultaneously and
independently a joint action, which triggers a transition to a
successor state, and generates a local payoff and possibly further
private signals to each player. 
Plays correspond to infinite paths through the graph and yield to each
player a global payoff according to a given aggregation function, such as mean payoff,
or parity. As solutions to such games, we are interested in
synthesising Nash equilibria in pure strategies, i.e.,
profiles of deterministic strategies that are self-enforcing 
when prescribed to all players by a central coordinator.

The basic setting is standard for the automated verification and
synthesis of reactive modules that maintain ongoing interaction with
their environment seeking to satisfy a common global specification.  
Generally, imperfect information about the play is
modelled as uncertainty about the current state in the underlying
transition system, whereas uncertainty about the actions of other players is
not represented explicitly. 
This is because the main question concerns \emph{distributed} winning
strategies, i.e., Nash equilibria in the special case where all
players receive maximal payoff, which here is common to all of them. 
If each player wins when all
follow the prescribed strategy, unilateral deviations cannot be
profitable 
and any reaction to them would be ineffective, 
hence there is no need to monitor actions of other players.
Accordingly, distributed winning strategies can be defined on
(potential) histories of visited states, independently of the history
of played actions. 
Nevertheless, these games are computationally
intractable in general, already with respect to the question of whether
distributed winning strategies exist~\cite{PnueliRos90,PetersonRei79,AzharPetRei01}.

Moreover, if no equilibria that yield maximal payoffs
exist for a given game, and we consider arbitrary Nash equilibria, 
% rather than distributed winning strategies, 
it becomes crucial for a player to monitor the actions of other
players. To illustrate,
one elementary scheme for constructing equilibria in games of
infinite duration relies on \emph{grim-trigger} strategies: cooperate
on the prescribed equilibrium path until one player deviates, and at
that event, enter a coalition with the remaining players and switch to a
joint punishment strategy against the deviator. 
Most procedures for
constructing Nash equilibria in games for verification and synthesis
are based on this scheme, which relies essentially on the ability of players 
to detect \emph{jointly} the deviation~\cite{Ummels08,UmmelsWoj11,Brenguier13,BrihayeBP14}. 
This works well under perfect, instant monitoring, where all players have
common knowledge about the last action performed by every other
player. 
However, the situation becomes more complicated 
when players receive only imperfect signals about the actions of other
players, and worse, if the signals are not delivered instantly, but
with uncertain delays that may be different for each player. 

To study the effect of imperfect, delayed monitoring on equilibria in
concurrent games, we introduce a refined model in which 
observations about actions are separated from observations about
states, and we incorporate a representation for nondeterministic
delays for observing action signals. 
To avoid the general undecidability results from the basic setting, 
we restrict to the case where the players have 
perfect information about the current state.
Under the assumption that the delays are uniformly bounded, we show
that equilibrium outcomes from the version of a game where signals are
delivered instantly
can be preserved in the variant where they are delayed. 
Towards this, we construct strategies for the
delayed-monitoring game by combining responses for the
instant-monitoring variant in such a way that any play with delayed
signals corresponds to a shuffle of several plays with instant
signals, which we call threads. 
Intuitively, delayed-monitoring strategies are constructed, in a
Frankenstein manner, from a
collection of instant-monitoring equilibrium strategies. 
Under an additional assumption that the payoff structure is
insensitive to shuffling plays this procedure allows to transfer
equilibrium values from the instant to the delayed-monitoring game.

We point out that when we set out with finite-state equilibrium strategies for the
instant-monitoring game, the procedure will also yield a profile of 
finite-state strategies for the delayed-monitoring game. Hence, the
construction is effective, and can be readily applied to cases 
where synthesis procedures for finite-state equilibria in games with
instant monitoring exist.  

\paragraph*{Related literature}

One motivation for studying infinite games with delays comes from the
work of Shmaya~\cite{Shmaya11} considering sequential games on finitely
branching trees (or equivalently, on finite graphs) 
where the actions of players are monitored
perfectly, but with arbitrary finite delays. In the setting of
two-player zero-sum games with Borel winning conditions, he shows that
these delayed-monitoring games are determined in mixed strategies. 
Apart of revealing that infinite games on finite graphs are robust
under monitoring delays, the paper is enlightening for its 
proof technique which relies on a reduction of the delayed-monitoring
game to a game with a different structure that features instant
monitoring but, in exchange, involves stochastic moves. 

Our analysis is inspired directly from a recent article of Fudenberg, Ishii, and
Kominers~\cite{FudenbergIK14} on infinitely repeated games
with bounded-delay monitoring whith stochastically
distributed observation lags. The authors prove a 
transfer result that is much stronger than ours, which also covers
the relevant case of discounted payoffs (modulo a controlled adjustment
of the discount factor). 
The key idea for constructing strategies in the delayed-response game
is to modify strategies from the instant-response game by letting them
respond with a delay equal to the maximal monitoring delay so that all
players received their signals. This amounts to 
combining different threads of the
instant-monitoring game, one for every time unit in the 
delay period. Thus, the proof again involves a reduction between games
of different structure, with the difference that here one game is
reduced to several instances of another one.

Infinitely repeated games correspond to the particular case of
concurrent games with only one state. This allows applying 
classical methods from strategic games which are no longer accessible
in games with several states~\cite{RosenbergSV06}.
Additionally, the state-transition structure of our setting 
induces a combinatorial effort to adapt the delayed-response 
strategies of Fudenberg, Ishii, and Kominers: As 
the play may reach a different state 
until the monitoring delay expires, 
the instant-monitoring threads must be scheduled more carefully
to make sure that they combine 
to a valid play of the delayed-monitoring variant. 
In particular, the time for returning to a particular game state may be
unbounded, which makes it hard to deliver guarantees under
discounted payoff functions. 
As a weaker notion of patience, suited for games with state
transitions, we consider payoff aggregation functions that are   
\emph{shift-invariant and submixing}, as introduced by Gimbert and Kelmendi
in their work on memoryless strategies in stochastic games~\cite{GimbertKel14}. 

Our model generalises 
concurrent games of infinite duration over finite graphs. Equilibria
in such models have been investigated for the perfect-information
case, and it was shown that it is decidable
with relatively low complexity whether equilibria exist, and
if this is the case, finite-state equilibrium profiles can be
synthesised for several relevant cases of 
interest~\cite{Ummels08,Ummels11,BouyerBMU11}.

The basic method for constructing equilibria in the 
case of perfect information relies on grim-trigger strategies that
react to deviations from the equilibrium path by turning to a
zero-sum coalition strategy opposing the deviating player. Such an approach
can hardly work under imperfect monitoring where deviating actions 
cannot be observed directly. 
Alternative approaches to constructing
equilibria without relying on perfect monitoring comprise, on the one hand
distributed winning strategies for games that allow all players of a
coalition to attain the most efficient
outcome~\cite{KupfermanVar01,FinkbeinerSch05,BKP11}, 
and at the other extreme, 
Doomsday equilibria, proposed by Chatterjee
et al. in~\cite{ChatterjeeEtAl14}, for games where any deviation
leads to the most inefficient outcome, for all players.

\medskip

In this paper, we prove a transfer result that implies effective
solvability of games with a particular kind of imperfect information,
due to imperfect monitoring of actions, and delayed delivery of
signals. 
Towards this, we first introduce a new model of concurrent games 
where observation signals associated to actions are detached from
the state information, and in which the emission and delivery time of
signals can be separated by a lag controlled by Nature. 
Then, we present the proof argument which relies on a reduction
of a delayed-monitoring game 
to a collection of instances of an instant-monitoring game.

\section{Games with delayed signals}

There are~$n$ players~$1, \dots, n$ and a distinguished agent called Nature.   
We refer to a list $x=(x^i)_{1\le i \le n}$ that associates one element $x^i$ to every
player~$i$ as a \emph{profile}.  
For any such profile, we write $x^{-i}$ to denote the 
list $( x^j )_{1 \le j \le n, j \neq i}$ where the element
of Player~$i$ is omitted. 
Given an element $x^i$ and a list $x^{-i}$, we denote by
$(x^i, x^{-i} )$ the full profile $(x^i)_{1 \le i  \le n}$. 
For clarity, we 
always use superscripts to specify to which player an element
belongs. If not quantified explicitly, we refer to Player~$i$
to mean any arbitrary player.

\subsection{General model}

For every player~$i$, we fix a set~$A^i$ of \emph{actions}, 
and a set~$Y^i$ of \emph{signals}; these sets are finite. 
The \emph{action space}~$A$ consists of all action profiles, and 
the \emph{signal space}~$Y$ of all signal profiles.

\subsubsection{Transition structure.} 
The transition structure of a game is described by 
a \emph{game graph} $G = (V, E)$ over a finite set~$V$ of
\emph{states} with
an edge relation 
$E \subseteq V \times A \times Y \times V$
that represents \emph{transitions} 
labelled by action and signal profiles. 
We assume that for each state $v$ and every action profile~$a$, 
there exists at least one transition $(v, a, y, v') \in E$. 

The game is played in stages over infinitely many periods 
starting from a designated initial state $v_0 \in V$ 
% that is chosen by Nature and 
known to all players. 
In each period~$t \ge 1$, starting in a state~$v_{t-1}$, 
every player~$i$ chooses an action $a^i_t$, 
and Nature chooses a transition
$(v_{t-1}, a_t, y_t, v_t) \in E$, which determines 
a profile~$y_t$ of emitted signals and a successor state $v_t$. 
Then, each player~$i$
observes a set of signals depending on the monitoring structure of the
game, 
and the play proceeds to period $t+1$ with 
$v_t$ as the new state.

Accordingly, a~\emph{play} is an infinite sequence 
$v_0,~ a_1, y_1, v_1,~ a_2, y_2, v_2 ~
\dots\in V(AYV)^\omega$ such that $(v_{t-1}, a_t, y_t, v_t) \in
E$, for all $t \ge 1$.
A \emph{history} is a finite prefix 
$v_0,~ a_1, y_1, v_1, ~\dots, ~a_t, y_t, v_t 
\in V (AYV)^*$
of a play.
We refer to the number of stages
played up to period~$t$ as the \emph{length} of the history.
%and we denote it by~$|\rho|$.

\subsubsection{Monitoring structure.} 
We assume that each player~$i$ always knows the current state $v$
and the action $a^i$ she is playing.
However, she is not informed about the actions or signals 
of the other players. Furthermore, she may observe 
the signal~$y^i_t$ emitted in a period $t$ only in some later period
or, possibly, never at all. 

The signals observed by Player~$i$ are described by an
\emph{observation} function 
\begin{align*}
  \beta^i:V (A Y V)^+ \to 2^{Y^i},
\end{align*} which
assigns to every nontrivial history 
$\pi = v_0,~ a_1, y_1, v_1, ~\dots, ~a_t, y_t, v_t$ 
with $t \ge 1$ a set 
of signals that were actually emitted along $\pi$ for the player:
\begin{align*}
  \beta^i( \pi ) \subseteq \{ y^i_r \in Y^i~|~ 1 \le r \le t \}.
\end{align*} 
For a global history~$\pi \in V(AYV)^*$, 
the \emph{observed history} of Player~$i$
is the sequence 
\begin{align*}
  \beta^i( \pi ) := v_0,~ a^i_1, z^i_1, v_1, ~  
  \dots, ~a^i_t, z^i_t, v_t
\end{align*} 
with 
$z^i_r = \beta^i( v_0, ~ a_1, y_1, v_1,~ \dots, ~ a_r, y_r, v_r )$, 
for all $1 \le r \le t$. Analogously, we define the
\emph{observed play} of Player~$i$.

A \emph{strategy} for player~$i$ is a mapping $s^i:V (A^i Y^i V)^* \to A^i$
that associates to every observation history $\pi \in V (A^i Y^i V)^*$ an action 
$s^i( \pi )$. The \emph{strategy space}~$S$ is the set of all strategy profiles.
We say that a history or a play $\pi$ \emph{follows} a strategy~$s^i$, if 
$a^i_{t+1} = s^i( \beta^i( \pi_t ))$, for 
all histories $\pi_t$ of length $t \ge 0$ in $\pi$. 
Likewise, a history or play 
follows a profile $s \in S$, if it follows the strategy $s^i$ of each
player~$i$.  
The \emph{outcome} $\mathrm{out}( s )$ of a strategy profile~$s$
is the set of all plays that follow it. 
Note that the outcome of a strategy profile 
generally consist of multiple plays,
due to the different choices of Nature.
 
Strategies may be partial functions. However, we require that for any
history~$\pi$ that follows a strategy $s^i$, the observation history 
$\beta^i( \pi )$ is also included in the domain of $s^i$.

With the above definition of a strategy, we implicitly assume that players have
perfect recall, that is, they may record all the information acquired
along a play. Nevertheless, in certain cases, we can restrict our
attention to strategy functions computable by automata with finite memory. 
In this case, we speak of \emph{finite-state strategies}.

\subsubsection{Payoff structure.} 
Every transition taken in a play
generates an integer payoff to each player~$i$,
described by a \emph{payoff} function $p^i: E \to \mathbb{Z}$. 
These stage payoffs are combined by a 
 \emph{payoff aggregation} function
$u: \mathbb{Z}^\omega \to \mathbb{R}$
to determine the \emph{utility} received by Player $i$ in a play~$\pi$ 
% $\pi = v_0,~ a_1, y_1, v_1,~ a_2, y_2, v_2 ~ \dots$
as 
$
  u^i( \pi ) := u (\, p^i( v_0, a_1, y_1, v_1), p^i(v_1, a_2, y_2,
v_2), \dots \,).
$
Thus, the profile of 
\emph{utility}, or global payoff, functions $u^i: V (A Y V)^\omega \to \mathbb{R}$ 
is represented by a profile of payoff functions~$p^i$ and 
an aggregation function~$u$, which is common to all players.

We generally consider utilities that
depend only on the observed play, that is, 
$u^i( \pi ) = u^i (\pi')$, for any plays $\pi, \pi'$ 
that are indistinguishable to Player~$i$, that is, $\beta^i( \pi ) =
\beta^i( \pi' )$.
To extend payoff functions from plays to strategy profiles, we set
\begin{align*}
  u^i( s ) := \inf \{ u^i( \pi ) ~|~ \pi \in \mathrm{out}( s ) \},
  ~\text{for each strategy profile } s \in S.
\end{align*}

Overall, a game $\mathcal{G} = (G, \beta, u)$ is 
described by a game graph with
a profile of observation functions and one of payoff functions. 
We are interested in \emph{Nash equilibria},
that is, strategy profiles $s \in S$ such that 
$u^i( s ) \ge u^i ( s^{-i}, r^{i} )$, for every player $i$
and every strategy $r^i \in S^i$. The payoff
$w = u^i( s )$ generated by an equilibrium~$s \in S$ 
is called an equilibrium payoff. An equilibrium payoff $w$ is
\emph{ergodic} if it does not depend on the initial state of the
game, that is, there exists a
strategy profile~$s$ with $u( s ) = w$, for every choice of an initial state.

\subsection{Instant and bounded-delay monitoring}

We focus on two particular monitoring structures, 
one where the players observe their component of the signal profile instantly, 
and one where each player~$i$ observes his private signal emitted in period $t$
in some period $t + d^i_t$, 
with a bounded delay~$d^i_t \in \mathbb{N}$ chosen by Nature.

Formally, a game with \emph{instant monitoring} is one where the 
observation functions~$\beta^i$ return, for every history
$\pi = v_0, ~ a_1, y_1, v_1, ~\dots, ~ a_t, y_t,v_t$ of length
$t\ge 1$,
the private signal emmited for Player~$i$ in the current stage, that is, 
$\beta^i(\pi) = 
\{ y^i_t \}$, for all $t \ge 1$. 
As the value is always a singleton, we may
leave out the enclosing set brackets and write $\beta^i( \pi ) = y^i_t$. 

To model bounded delays, we consider signals
with an additional component that represents a timestamp. 
Concretely, we fix a set $B^i$ of \emph{basic signals} 
and a finite set $D^i \subseteq \mathbb{N}$ of
\emph{possible delays}, for each player~$i$, and consider the product 
$Y^i := B^i \times D^i$ as a new set of signals. 
Then, a game with \emph{delayed monitoring} is a game over the signal space $Y$ 
with observation functions~$\beta^i$ that return, 
for every history $\pi = v_0, ~ a_1,
(b_1,d_1), v_1, ~\dots, ~ a_t, (b_t, d_t), v_t$ of length~$t
\ge 1$, the value 
\begin{align*}
  \beta^i( \pi ) = \{ (b^i_r, d^i_r) \in B^i\times D^i~|~r \ge 1,~r + d^i_r =
  t \}. 
\end{align*}

In our model, the role of Nature 
is limited to choosing the delays for
observing the emitted signals.
Concretely, we postulate that the basic signals 
and the stage payoffs associated to transitions
are determined by the current state and the action profile chosen by
the players, that is, for every global state $v$ and action profile
$a$, there exists a unique profile $b$ of basic signals and a unique 
state~$v'$ such that $(v, a, (b, d), v') \in E$, for some $d \in
D$; moreover, for any other delay profile $d' \in D$, 
we require $(v, a, (b, d'), v' ) \in E$, and also that 
$p^i(v, a, (b, d), v' )= p^i(v, a, (b, d'), v' )$.
Here again, $D$ denotes the \emph{delay space} 
composed of the sets~$D^i$. Notice that under this assumption, the
plays in the outcome of a strategy profile $s$ differ only by
the value of the delays. In particular, all plays in $\mathrm{out}( s )$
yield the same payoff.

To investigate the effect of observation delays, 
we will relate the delayed and instant-monitoring 
variants of a game. 
Given a game $\mathcal{G}$ with delayed monitoring, 
the corresponding instant-monitoring game $\mathcal{G'}$ is
obtained by projecting every signal $y^i = (b^i, d^i)$
onto its first component $b^i$ and 
then taking the transition and payoff structure induced by
this projection. As we assume that 
transitions and payoffs are independent of delays, 
the operation is well defined. 

Conversely, given a game $\mathcal{G}$ with instant monitoring 
and a delay space~$D$, 
the corresponding game $\mathcal{G}'$ with delayed monitoring is obtained by
extending the set $B^i$ of basic signals in $\mathcal{G}$
to $B^i \times D^i$, for each player~$i$, and
by lifting the transition and payoff structure accordingly. Thus, the
game $\mathcal{G}'$  has the same states as $\mathcal{G}$ 
with transitions $E' := \{ (v, a, (b,d), w)~|~ (v, a, b, w) \in E, d \in D \}$, 
whereas the payoff functions are given 
by ${p'}^i( v, a, (b, d), w) := p^i( v, a, b, w )$, 
for all~$d \in D$.

As the monitoring structure of games with instant or delayed monitoring
is fixed, it is sufficient to describe  
the game graph together with the profile of payoff functions, and to
indicate the payoff aggregation function. It
will be convenient to include the payoff associated to a transition as an
additional edge label and thus represent the game simply as a pair
$\mathcal{G} = (G, u)$ consisting of a finite labelled game graph and an 
aggregation function $u: \mathbb{Z}^\omega \to \mathbb{R}$.

\subsection{Shift-invariant, submixing utilities}

Our result applies to a class of games where the payoff-aggregation
functions are invariant under removal of prefix histories and 
shuffling of plays. Gimbert and Kelmendi \cite{GimbertKel14}
identify these properties as a guarantee for the existence of simple
strategies in stochastic zero-sum games.

A function $f: \mathbb{Z}^\omega \to \mathbb{R}$ is \emph{shift-invariant},
if its value does not change when adding an arbitrary finite prefix to the
argument, that is, for every sequence $\alpha \in \mathbb{Z}^\omega$ 
and each element $a \in \mathbb{Z}$, we have 
$f( a \alpha ) = f( \alpha )$.

An infinite sequence $\alpha \in \mathbb{Z}^\omega$ is a \emph{shuffle} 
of two sequences $\beta, \gamma \in \mathbb{Z}^\omega$, 
if $\mathbb{N}$ can be partitioned into two infinite sets 
$I = \{ i_0, i_1, \dots \} $ and $J = \{j_0, j_1, \dots \}$ such that 
$\alpha_{i_k} = \beta_k$ and $\alpha_{j_k} = \gamma_k$, for all $k \in
\mathbb{N}$. 
A function $f: \mathbb{Z}^\omega \to \mathbb{R}$ is called \emph{submixing} 
if, for every shuffle $\alpha$ of two sequences $\beta, \gamma \in
\mathbb{Z}^\omega$, we have 
\begin{align*}
\min\{ f( \beta ), f( \gamma) \} \le f( \alpha ) \le \max \{ f(
\beta), f( \gamma )\}.
\end{align*}
In other words, the image of a shuffle product always lies between
the images of its factors.

The proof of our theorem relies on payoff aggregation functions
$u : \mathbb{Z}^\omega \to \mathbb{R}$ that are shift-invariant and 
submixing. Many relevant game models used in
economics, game theory, and computer science satisfy this restriction.
Prominent examples are mean payoff or limsup payoff, 
which aggregate sequences of stage payoffs $p_1, p_2,
\dots\in \mathbb{Z}^\omega$ by setting:
\begin{align*}
  \mean ( p_1, p_2, \dots) & := \limsup_{t \ge 1} \frac{1}{t} \sum_{r =
    1}^{t} p_r, \quad \text{and} \\
  \LimSup( p_1, p_2, \dots) & := \limsup_{t \ge 1} p_t. 
\end{align*}

Finally, parity conditions which map non-negative integer payoffs 
$p_1, p_2, \dots $ called priorities  to
$\parity(p_1, p_2, \dots ) = 1$ if the least priority that occurs 
infinitely often is even, and $0$ otherwise, also satisfy the conditions.

\subsection{The transfer theorem}

We are now ready to formulate our result stating that, under certain restrictions,
equilibrium profiles from games with instant monitoring can be
transferred to games with delayed monitoring. 

\begin{theorem}
Let $\mathcal{G}$ be a game with instant monitoring and
shift-invariant submixing payoffs, and let $D$ be a finite delay
space~$D$. Then, for every ergodic equilibrium payoff $w$ in $\mathcal{G}$, 
there exists an equilibrium of the $D$-delayed monitoring game
$\mathcal{G}'$ with the same payoff~$w$. 
\end{theorem}

The proof relies on constructing a strategy for the delayed-monitoring
game while maintaining a collection of virtual plays of the
instant-monitoring game on which the given strategy is queried. 
The responses are then combined according to a
specific schedule to ensure that the actual play arises as a shuffle
of the virtual plays.

\section{Proof}

Consider a game $\mathcal{G} = (G, u)$ with instant monitoring where
the payoff aggregation function $u$ is shift-invariant and submixing,
and suppose that $\mathcal{G}$ admits an equilibrium profile $s$. 
For an arbitrary finite delay space $D$, let 
$\mathcal{G'}$ be the delayed-monitoring variant of $\mathcal{G}$.
In the following steps, we will construct a strategy profile $s'$ for $\mathcal{G'}$, that is in
equilibrium and yields the same payoff $u( s' )$ as $s$ in $\mathcal{G'}$.

\subsection{Unravelling small cycles}

To minimise the combinatorial overhead for scheduling delayed responses,
it is convenient to ensure that, 
whenever the play returns to a state $v$, the signals emitted at the
previous visit at $v$ have been received by all players. 
If every cycle in the given game graph~$G$ is at least as long as any possible
delay, this is clearly satisfied. Otherwise,  
the graph can be expanded to avoid small cycles, e.g., by
taking the product with a cyclic group of order equal to the 
maximal delay. 

Concretely, let $m$ be the greatest delay among $\max D^i$, for all players $i$. 
We define a new game graph $\hat{G}$ as the product of~$G$ 
with the additive group $\mathbb{Z}_m$ of integers modulo~$m$, 
over the state set $\{ v_j ~|~ v \in V, j \in \mathbb{Z}_m \}$ 
by allowing transitions $(v_j, a, b, v'_{j+1})$, 
for every $(v, a, b, v') \in E$ and all $j \in
 \mathbb{Z}_m$, and by assigning stage payoffs 
$\hat{p}^i(v_j, a, b, v'_{j+1}) := p^i(v, a, b, v')$, 
for all transitions $(v, a, b, v') \in E$. Obviously,
every cycle in this game has length at least $m$. 
Moreover, the games $(\hat{G}, u)$ and $(G, u )$ are equivalent: Since the
index component $j \in \mathbb{Z}_m$ is not observable to the players, 
the two games have the same sets of strategies, 
and profiles of corresponding strategies yield the same observable
play outcome, and hence the same payoffs.

In conclusion, we can assume without loss of generality 
that each cycle in the game graph~$G$ is longer than the maximal
delay $\max D^i$, for all players $i$.

\subsection{The Frankenstein procedure}

We describe a strategy $f^i$ for Player~$i$ in the delayed monitoring game
$G'$ by a reactive procedure that receives observations of states and
signals as input and produces actions as output. 

The procedure 
maintains a collection of virtual plays of the
instant-monitoring game. More precisely, these are observation
histories for Player~$i$ following the strategy $s^i$ in~$G$, 
which we call \emph{threads}. The observations collected
in a thread $\pi = v_0,~a_1^i, (b_1^i, d_1^i), v_1, ~\dots,~
a_r^i, (b_r^i, d_r^i), v_r$ are drawn from the play of 
the main delayed-monitoring game~$G'$. Due to delays, 
it may occur that the signal $(b_r^i, d_r^i)$ emitted in the last
period of a thread has not yet been received. 
In this case, the signal entry is replaced by a special symbol
$\#$, and we say that the thread is \emph{pending}. 
As soon as the player
receives the signal, the placeholder $\#$ is overwritten with the actual value, and the
thread becomes \emph{active}. Active threads $\pi$ are used to query the 
strategy~$s^i$; the prescribed action $a^i = s^i ( \pi )$ is
played in the main delayed-monitoring game and it is also used to continue
the thread of the virtual instant-monitoring game.
 
To be continued, a thread must be active and
its current state needs to match the
actual state of the play in the delayed-monitoring game. Intuitively,
threads advance more slowly than the actual play, so we need
multiple threads to keep pace with it. 
Here, we use a collection of $|V|+1$ threads, indexed by an 
ordered set $K = V \cup \{ \varepsilon \}$. 
The main task of the procedure is to schedule the continuation of
threads. To do so, it
maintains a data structure $(\tau, h)$ that
consists of the threads $\tau = (\tau_k)_{k \in K}$ and
a scheduling sequence $h = h[0], \dots, h[t]$  of indices from $K$, 
at every period $t \ge 0$ of the actual play. 
For each previous $r < t$, the entry $h[r]$ points to the thread
according to which the action of period $r+1$ in the actual play has
been prescribed; the
last entry $h[t]$ points to an active thread that is currently scheduled for
prescribing the action to be played next.

The version of Procedure Frankenstein${}^i$ for Player~$i$, given
below, is para\-metrised by the game graph $G$ with the designated initial state, the
delay space $D^i$, and the given equilibrium strategy $s^i$ in the
instant-monitoring game.  In the initialisation phase, the
initial state $v_0$ is stored in the initial thread
$\tau_\varepsilon$ to which the current scheduling entry $h[0]$ points. 
The remaining threads are
initialised, each with a different position from $V$.
Then, the procedure enters a non-terminating loop along the periods
of the actual play. 
In every period~$t$, it outputs
the action prescribed by strategy $s^i$ for the current thread scheduled by
$h[t]$ (Line 5). 
Upon receiving the new state, this current thread 
is updated by recording the played action and the successor
state; as the signal emmitted in the instant-monitoring play is not
available in the delayed-monitoring variant, it is temporarility replaced by
$\#$, which marks the current thread as pending (Line 7). 
Next, an active thread that matches the new state is scheduled (Line
9), and the received signals are recorded with the pending threads 
to which they belong (Line 11 -- 14). As a consequence, these threads become
active.

\setlength{\interspacetitleruled}{0pt}%
\setlength{\algotitleheightrule}{0pt}%

\SetKw{KwAssert}{assert}
\SetKw{KwPlay}{play}
\SetKw{KwReceive}{receive}
\begin{algorithm}

\smallskip
\TitleOfAlgo{Frankenstein${}^{i}(G, v_0, D^i, s^i)$}
% \caption{Frankenstein}
\label{fig:frankenstein}
\tcp{initialisation}
\nl
$\tau_{\varepsilon} := v_0$; $h[0] = \varepsilon$\;
\nl
\lForEach {$v \in V$} {$\tau_{v} := v$}\;
\medskip

\tcp{play loop }

\For {$t = 0$ \KwTo $\omega$ } {
\nl
$k := h[t]$\;
\nl
\KwAssert ($\tau_k$ is an active thread)\;
\nl
\KwPlay action $a^i:= s^i( \tau_k )$  \tcp*[r]{$a_{t+1}^i$ \hspace*{5cm}}
\nl
\KwReceive new state $v$  \tcp*[r]{$v_{t+1}$\hspace*{5.2cm}}
\nl
update $\tau_k := \tau_k \, a^i \# v$\;
\medskip
\nl
\KwAssert (there exists an index $k' \neq k$ such that 
$\tau_{k'}$ ends at state $v$)\;
\nl
set $h[t+1]$ to the least such index $k'$ \;
\medskip
\nl
\KwReceive observation $z^i \in B^i \times D^i$
\tcp*[r]{$z_{t+1}^i$\hspace*{4cm}}
\nl
\ForEach{$(b^i, d^i) \in z^i$ } {
  \nl
  $k := h[t-d^i]$\;
  \nl
  \KwAssert ($\tau_k = \rho \# v'$, for some prefix $\rho$, state
  $v'$)\; 
  \nl
  update $\tau_k := \rho (b^i, d^i) v'$\;
}
}

\label{algo:frankenstein}

\end{algorithm}

\subsection{Correctness}

In the following, we argue that the procedure Frankenstein${}^i$ never violates the
assertions in Line 4, 8, and 13 while interacting with Nature in the
delayed-monitoring game~$\mathcal{G'}$, and thus implements a valid strategy for
Player~$i$. 

Specifically, we show that for every history
\begin{align*}
  \pi = v_0,~a_1, (b_1, d_1), v_1,~\dots,~a_{t}, (b_{t}, d_{t}),
v_{t}
\end{align*} 
in the delayed-monitoring game that follows the prescriptions
of the procedure up to period $t>0$, (1) the
scheduling function $h[t] = k$ points to an active thread~$\tau_k$ 
that ends at state~$v_t$, and (2) for 
the state~$v_{t+1}$ reached by playing $a_{t+1} := s^i( \tau_k )$ at~$\pi$, 
there exists an active thread $\tau_{k'}$ that ends at $v_{t+1}$. 
We proceed by induction over the period~$t$. In the base case,
both properties hold, due to the way in which the data structure is
initialised: the (trivial) thread $\tau_\varepsilon$ is active, and
for any successor state $v_1$ reached by $a_1 := s^i(
\tau_\varepsilon)$, 
there is a fresh thread $\tau_{v_1}$ that is active. For the induction
step in period $t+1$, property (1) follows from property (2) of
period~$t$. 
To verify that property (2) holds, we distingiush two
cases. If $v_{t+1}$ did not occur previously in $\pi$, the initial
thread $\tau_{v_{t+1}}$ still consists of the trivial history
$v_{t+1}$, and it is thus active. 
Else, let $r < t$ be the period in which $v_{t+1}$ occurred
last. Then, for $k' = h[r]$, the thread $\tau_{k'}$ ends at
$v_{t+1}$. Moreover, 
by our assumption that the cycles in $G$ are longer than any
possible delay, it follows that the signals emitted in period 
$r < t - m$ have been received along $\pi$ and were recorded (Line
12--14). Hence, $\tau_{k'}$ is an active thread ending at $v_{t+1}$,
as required. 

To see that the assertion of Line 13 is never violated, 
we note that every observation history $\beta^i(\pi)$ of the actual
play $\pi$ in $\mathcal{G'}$ up to period $t$
corresponds to a finitary shuffle of the threads $\tau$ in the $t$-th iteration
of the play loop, described by the scheduling function $h$: The
observations 
$(a_r^i,(b_r, d_r)^i, v_r)$ associated to any period $r \le t$
appear at the end of $\tau_{h[r]}$, if the signal $(b_r, d_r)^i$ was
delivered until period $t$, and with the placeholder $\#$, otherwise.

In summary, it follows that 
the reactive procedure Frankenstein${}^i$ never halts, and 
it returns an
action for every observation $\beta^i( \pi )$ of a history
$\pi$ that follows it. Thus, the procedure defines a strategy
$f^i: V (A^iY^iV)^* \to A^i$  for Player~$i$.

\subsection{Equilibrium condition}

Finally, we show that the interplay of the strategies $f^i$ described by the
reactive procedure Frankenstein${}^i$, for each player~$i$,
constitutes an equilibrium profile for the delayed-monitoring game~$\mathcal{G'}$ 
yielding the same payoff as $s$ in $\mathcal{G}$. 

According to our remark in the previous subsection, every transition
taken in a play $\pi$ that follows the strategy $f^i$ in $\mathcal{G'}$  
is also observed in some thread history, which in turn follows $s^i$. 
Along the non-terminating execution
of the reactive Frankenstein${}^i$ procedure, some threads must be scheduled
infinitely often, and thus correspond to observations of plays in the
perfect-monitoring game~$G$. We argue that the observation
by Player~$i$
of a play that follows the
strategy~$f^i$ corresponds 
to a shuffle of such infinite
threads (after discarding finite prefixes).

To make this more precise, let us fix a play $\pi$ that follows $f^i$
in~$\mathcal{G'}$, and consider the infinite
scheduling sequence $h[0], h[1], \dots$ generated by the procedure. 
Since there are finitely many
thread indices, some must appear infinitely often in this sequence; we
denote by $L^i \subseteq K$ the subset of these indices, and look at the
least period $\ell^i$, after which only threads in $L^i$ are
scheduled. Then, the suffix of the observation $\beta^i(\pi)$ from
period $\ell^i$ onwards can be written as a $|L^i|$-partite shuffle 
of suffixes of the threads $\tau_k$ for $k \in L^i$. 

By our assumption that the payoff aggregation function $u$ is
shift-invariant and submixing, it follows that the payoff $u^i( \pi )$
lies between $\min \{  u^i (\tau_k)~|~k \in L^i \}$ and $\max \{ u^i
(\tau_k) ~|~ k \in L \}$.
Now, we apply this reasoning to all players to 
show that $f^i$ is an equilibrium profile with
payoff $u( s )$. 

To see that the profile $f$ in the delayed-monitoring game~$\mathcal{G'}$ 
yields the same payoff as $s$ in the instant-monitoring
game~$\mathcal{G}$, 
consider the unique play $\pi$ that
follows $f$, and construct $L^i$, for all players~$i$, as above. Then,
all threads of all players~$i$ follow~$s^i$, which by ergodicity implies, 
for each infinite thread $\tau_k$ with $k \in L^i$ that $u^i( \tau_k )
= u^i( s )$. Hence $\min \{  u^i (\tau_k)~|~k \in L^i \}$ = $\max \{ u^i
(\tau_k) ~|~ k \in L \}$= $u^i( \pi )$, for each player~$i$, and therefore
$u( f ) = u( s )$. 

To verify that $f$ is indeed an equilibrium profile, 
consider a strategy $g^i$ for the delayed-monitoring game and look at the 
unique play~$\pi$ that follows $(f^{-i}, g^i)$ in $\mathcal{G'}$. 
Towards a contradiction, assume that $u^i ( \pi ) > u^i ( f )$. Since 
$u^i ( \pi ) < \max \{ u^i (\tau_k) ~|~ k \in L^i \}$, there must exist
an infinite thread $\tau_k$ with index $k \in L^i$ such that 
$u^i( \tau_k ) > u^i( f ) = u^i( s )$. But $\tau_k$ corresponds to the
observation $\beta^i( \rho )$ of 
a play $\rho$ that follows $s^{-i}$ in $\mathcal{G}$, and since 
$s$ is in equilibrium strategy we obtain $u^i( s ) \ge u^i( \rho ) =
u^i( \tau_k)$, a contradiction. 

This concludes the proof of our theorem.

\subsection{Finite-state strategies}

The transfer theorem makes no assumption on the complexity of
equilibrium strategies in the instant-monitoring game at the
outset; informally, we may think of these strategies as 
oracles that the Frankenstein procedure can query. 
Moreover, the procedure itself runs for infinite time along the
periods the play, and the data structure it maintains grows
unboundedly. 

However, if we set out with an equilibrium profile of finite-state
strategies, it is straightforward to rewrite the Frankenstein
procedure as a finite-state automaton: instead of storing the
the full histories of threads, it is sufficient to maintain the curren
state reached by the the strategy automaton for the relevant player
after reading this history, over a period that is sufficiently long to
cover all possible delays.

\begin{corollary}
Let $\mathcal{G}$ be a game with instant monitoring and
shift-invariant submixing payoffs, and let $D$ be a finite delay
space~$D$. Then, for every ergodic payoff $w$ in
$\mathcal{G}$ generated by a profile of finite-state strategies, 
there exists an equilibrium of the $D$-delayed monitoring game
$\mathcal{G}'$ with the same payoff~$w$ that is also generated by a
profile of finite-state strategies. 
\end{corollary}

{
\bibliographystyle{siam}
\bibliography{delays}
}

\end{document}